\documentclass[letterpaper, 10 pt, conference]{ieeeconf}  % Comment this line out if you need a4paper
\IEEEoverridecommandlockouts        
\usepackage{graphicx}
\usepackage{amsmath} % assumes amsmath package installed
\usepackage{amssymb}  % assumes amsmath package installed
\usepackage{comment}

\title{\LARGE \bf
Predicting Trust Dynamics with Dynamic SEM in Human-AI Cooperation
}
\author{Sota Kaneko$^{1,2}$ and Seiji Yamada$^{2,1}$% <-this % stops a space
\thanks{*This work was supported in part by JST CREST JPMJCR21D4.}% <-this % stops a space
\thanks{$^{1}$Department of Informatics, 
        The Graduate University for Advanced Studies, SOKENDAI,
        Tokyo, Japan
        {\tt\small sota@nii.ac.jp}}%
\thanks{$^{2}$Digital Content and Media Sciences Research Division,
        National Institute of Informatics,
        Tokyo, Japan
        }%
}
\begin{document}
\maketitle
\thispagestyle{empty}
\pagestyle{empty}
\begin{abstract}
Humans' trust in AI constitutes a pivotal element in fostering a synergistic relationship between humans and AI. This is particularly significant in the context of systems that leverage AI technology, such as autonomous driving systems and human-robot interaction. Trust facilitates appropriate utilization of these systems, thereby optimizing their potential benefits. If humans over-trust or under-trust an AI, serious problems such as misuse and accidents occur. To prevent over/under-trust, it is necessary to predict trust dynamics. However, trust is an internal state of humans and hard to directly observe. Therefore, we propose a prediction model for trust dynamics using dynamic structure equation modeling, which extends SEM that can handle time-series data. A path diagram, which shows causalities between variables, is developed in an exploratory way and the resultant path diagram is optimized for effective path structures. Over/under-trust was predicted with 90\% accuracy in a drone simulator task,, and it was predicted with 99\% accuracy in an autonomous driving task. These results show that our proposed method outperformed the conventional method including an auto regression family. 
\end{abstract}
\section{Introduction}

AI technologies have been developed in various fields, and its use in everyday situations, such as autonomous driving, autonomous flying drones, and autonomous mobile robots, is rapidly advancing.
The development of such technology allows people to delegate tasks to them, reducing their workload.
While there are cases where all operations are simply left to autonomous driving or autonomous mobile robots, in many cases, they work together with humans in the same space.
In this way, appropriate cooperation between humans and AI is indispensable in the use and development of AI technology, and what becomes important here is trust of humans in AI~\cite{okamura2020adaptive,okamura2020emprical}.

When humans overestimate the performance of AI beyond its actual capabilities, there is a risk of misuse, such as delegating tasks in situations where they should not be delegated.
For example, in the case of autonomous driving, continuing to drive automatically despite a decrease in AI performance due to worsening weather conditions can lead to accidents.
This overestimation of AIs performance is referred to as over-trust\cite{Visser2020Calibration}.
On the other hand, underestimating the performance of AI excessively compared with its original capabilities can result in humans performing tasks that AI can carry out, preventing AI from demonstrating its actual performance.
This under-trust, or excessive underestimation, also presents a problem of decreased efficiency in use.
Therefore, maintaining appropriate trust in AI is important for proper collaboration with AI.

Also, in situations where humans and AI collaborate to perform tasks, predicting trust dynamics becomes important.
In real-time systems, such as those represented by autonomous driving, the performance of AI and the trust in AI, which is a human’s estimate of AIs performance, continue to change as the surrounding situation changes over time.
This change in trust is called trust dynamics, and if we can predict trust dynamics, we can also predict over-trust and under-trust.
This allows us to prevent falling into over-trust and under-trust before it happens.
However, trust in AI is an internal state of humans, so it is impossible to observe it directly from the outside.
Therefore, to predict trust dynamics, it is necessary to deal with this latent value.

Therefore, we construct a prediction model for trust dynamics, which is a change in the internal state of humans called "trust". In this work, we apply dynamic structural equation modeling (DSEM) for modeling trust and predicting over/under trust in a efficient and explainable way~\cite{Molnar2023interpretable}. We consider this work is the original approach which try to predict trust dynamics including direct prediction of over/under-trust by using DSEM. The procedures to construct a prediction model consist of exploratory design of path structure and its simple optimization for the most effective structure. We conducted experiments to evaluate our proposed methods in vision-based object recognition and autonomous driving simulation, and obtained promising results.  

%%%%%%%%%%%%%%%%%%%%
% 提案手法の説明を追加
%%%%%%%%%%%%%%%%%%%%

%%%%%%%%%%%%%%%%%%%%%%%%%%%%%%%%%%%%%%%%%%%%%%%%%%%%%%%%%%%%%%%%%%%%%%%%%%%%%%%%
\section{Related Work}
\subsection{Trust in HRI}
In research on trust formation in human-robot interaction (HRI), the factors that influence trust are classified as follows~\cite{Baker2018TrustInHRI,Khavas2020survey}: factors related to the robot (agent), factors related to the task and environment, and factors related to humans. 
%\begin{itemize}
%    \item Factors related to the robot (agent)
%    \item Factors related to the task and environment
%    \item Factors related to humans
%\end{itemize}
The impact on trust formation is greatest in the order of factors related to the robot, factors related to the task and environment, and factors related to humans.
Factors related to the performance of the robot include the reliability of the robot, the timing and frequency of task failures, the transparency of the system, etc., which are considered to determine the quality of the robots operation.
Factors related to the task and environment include rationality, the danger that the task poses to humans, the load and complexity of the task, etc. Factors related to humans include personality, knowledge about the system, past experiences with robots, etc.
Also, in HAI, trust formation can be considered in the same way by replacing factors related to the robot with factors related to the agent, but it is necessary to consider the difference resulting from the presence or absence of a physical entity.

\subsection{Trust Dynamics}
In interactions between humans and agents, trust changes over time and with repeated interactions. This changing trust is referred to as trust dynamics. By capturing trust dynamics, it is also possible to more accurately understand the major factors that influence trust formation.

In Luo's study on trust dynamics in interactions with autonomous vehicles, it is shown that changes in performance due to internal system factors have a greater impact on trust than changes in performance due to external system factors~\cite{Luo2020Human-AV}.
Performance degradation due to internal system factors is caused by things like sensor failures, while performance degradation caused by external factors includes detours due to road construction and increased travel time due to traffic congestion.

Furthermore, it has been shown that by incorporating the mechanism of trust, it is possible to make more accurate trust predictions than models that do not consider these factors~\cite{Collins2016Cognitive}.
In addition, there are efforts to develop new modeling methods that use trust dynamics, such as improving the accuracy of trust prediction by clustering based on trust dynamics and forming a suitable trust prediction model for each cluster~\cite{Liu2021Clustering}.

Thus, constructing a model that takes into account trust dynamics not only enables accurate prediction of trust, but also enables accurate understanding of the factors that influence trust.

\subsection{Determination of Over/Under-trust by Equation of Inequality}
In Lee study on designing reliance, over-trust is poor calibration in which trust exceeds system capability, and under-trust is trust falls of the system capability~\cite{Lee2004Designing}.

Okamura~\cite{okamura2020adaptive,okamura2020emprical} proposed a framework for determining over/under-trust from a \emph{reliance equation} involving the relationship between AI performance and human performance. In human-AI cooperative decision-making task that involving image recognition on a drone simulator, over/under-trust is determined from actually monitoring human rational decision-making behaviors based-on the reliance equation. The reliance equation is described like $T^H_A >=< T_A$, where $T^H_A$ and $T_A$ are AI's task success probabilities estimated by a human and the true value.

In this formulation, since over/under-trust can be detected only when a sequence of human decision-making behaviors by over/under-trust, it is hard to predict over/under-trust and prevent it. In contrast, our proposed method can predict over/under-trust and prevent it in advance. 

% In their experiment, overconfidence predictions were made in a human-AI collaborative decision-making task involving image classification of captured images by drone.

%In this framework, the AIs task success probability, which is the AIs performance, is denoted as \(P_{auto}\), \(P_{auto}\) estimated by humans is denoted as \(P_{trust}\), and the humans task success probability, which is the humans performance, is denoted as \(P_{auto}\). The following inequalities were defined to express over-trust and under-trust. \(\hat{P}_{man}\) is the humans estimate of human performance by the human themselves.
%    \begin{equation}
%        (P_{trust} > \hat{P}_{man}) \land (P_{man} > P_{auto})
%        \label{eq:okamura-ot}
%    \end{equation}
%    \begin{equation}
%        (P_{trust} < \hat{P}_{man}) \land (P_{man} < P_{auto})
%        \label{eq:okamura-ut}
%    \end{equation}
%Equation \ref{eq:okamura-ot} indicates an over-trust state, indicating that despite the AI performance being lower than the humans performance, the human estimates that they are inferior to the AI. Equation \ref{eq:okamura-ut} indicates a state of under-trust, indicating that despite the AIs performance being higher than the humans performance, the human estimates that they are superior to the AI. Since \(P_{trust} > \hat{P}{man}\) and \(P{trust} < \hat{P}_{man}\) can be measured by human behavior as to whether or not a task is entrusted to an AI, over-trust is determined by combining human behavior with the size relationship between human performance and AI performance.

\subsection{Trust Prediction}
Fukuchi used a Transformer to predict reliance and performed reliance calibration using a reliance calibration queue~\cite{fukuchi2023selectively}.
It should be noted that the reliance used here has a different definition from the trust we use.

Xu used a dynamic Bayesian network to predict trust dynamics~\cite{Xu2015OPTIMo}.
A cooperative decision-making task was performed using a drone simulator, where humans intervene in the automatic control of the drone, and trust dynamics were predicted in situations where the environment changed over time. 
The Bayesian network was constructed using six variables: trust, AI performance, presence or absence of user intervention, changes in external factors, changes in trust, and feedback on trust. Trust at time \(T = k\) is predicted from observations at times \(T = k, k-1\).

%%%%%%%%%%%%%%%%%%%%%%%%%%%%%%%%%%%%%%%%%%%%%%%%%%%%%%%%%%%%%%%%%%%%%%%%%%%%%%%%
\section{Method}
\subsection{Algorithm for Dynamic Path Diagrams}\label{sec:algo4path}
To predict and construct an explainable model of trust dynamics, which is an internal state that cannot be directly observed from the outside, we use DSEM, which extends structural equation modeling (SEM) to time series data.
SEM is a model that estimates causal relationships by performing path analysis between variables, dealing with observed variables that can be directly observed and measured, and latent variables that cannot be measured~\cite{kline2023principles}.
This SEM is extended along the time axis for the purpose of handling time series data and predicting the value at the next time point, which is DSEM~\cite{Asparouhov2018DSEM}.

By using DSEM as a prediction model, there are mainly three advantages:
\begin{itemize}
    \item It is possible to handle the concept of trust, which is defined as the value of AIs performance estimated by humans and is an internal state of humans, as a variable.
    \item It is possible to handle time series data and predict trust at the next time step.
    \item the edges (paths) spanning between nodes are given top-down on the basis of prior research, so the model has high interpretability.
\end{itemize}

Our proposed method of constructing a prediction model of trust dynamics can be summarized in the following steps.

\begin{enumerate}
    \item Exploratory design of path diagrams: A human designs an initial static path diagram for SEM based on domain knowledge containing insights from previous work and designer knowledge, and it is improved until the accuracy reaches a threshold $\tau$. This is done with human-in-the-loop procedures. \label{step-1}
    \item Optimization of time-series structure: Dynamic path diagrams based on the static path diagram (Step~\ref{step-1}) are automatically optimized with edges manually added between the path diagrams with different time steps. Optimization can be done by using a constrained-brute-force search algorithm which searches for all candidates of partial sequences within a constrained time range $\eta$. The objective function is time-series rolling-origin cross-validation~\cite{Tashman2000}. Since the computational complexity of this search is $O(2^n)$, the exponential order of time-series length $n$ is extremely large, so we introduced a hyper parameter, that it, the constrained time range $\eta$ which can be heuristically set.
\end{enumerate}

\subsection{Construction of Dynamic Path Diagrams}
\begin{figure}[tbp]
    \centering
    \includegraphics[width=86mm]{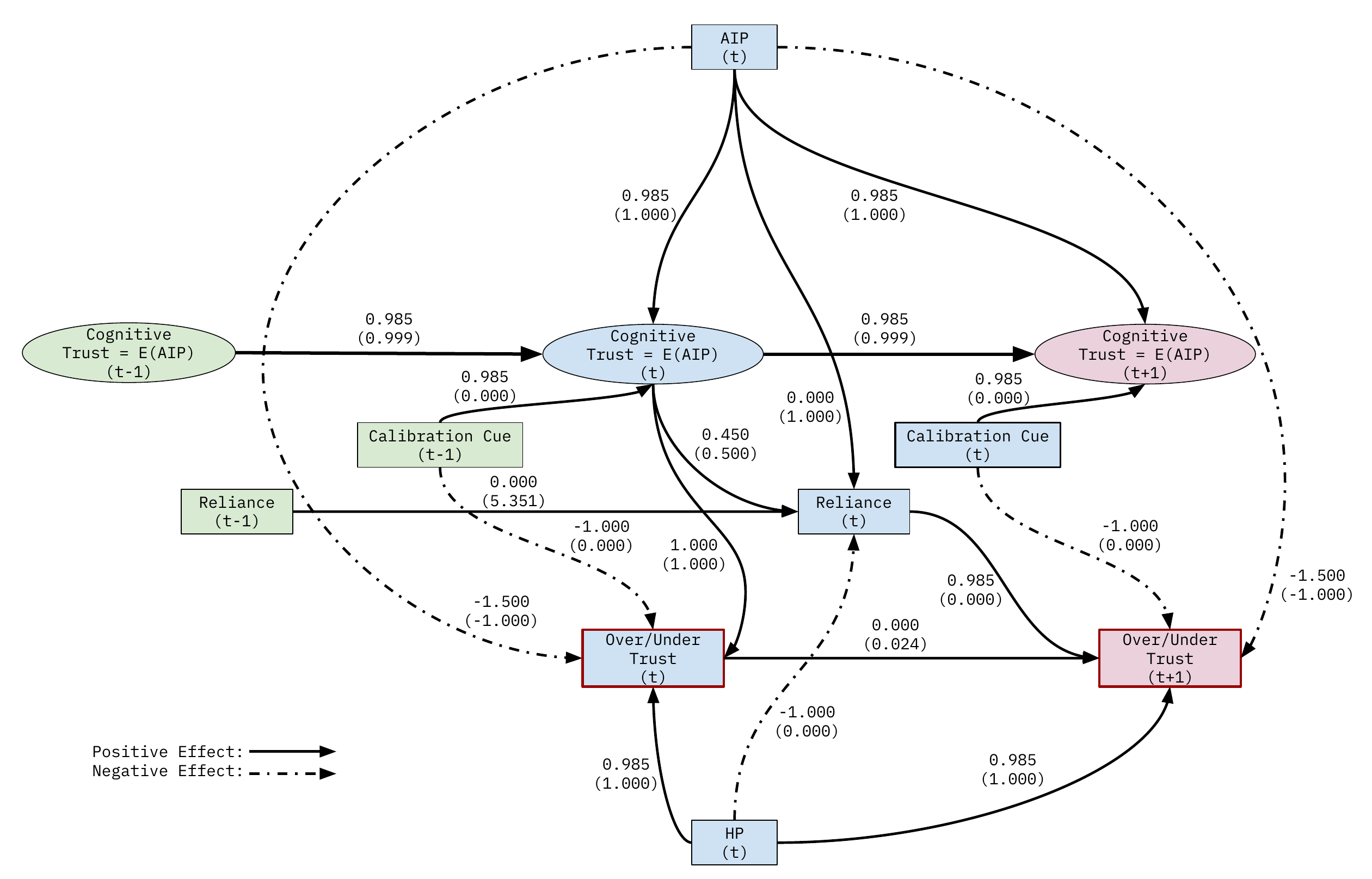} % width=160mm -> 86mm
    \caption{Dynamic path diagram with path coefficients in Exp-1 and Exp-2.}
    \label{fig:sem-after}
\end{figure}

%% D-SEM Model
On the basis of previous research, a model was created by Step~\ref{step-1} of Section III A with $\tau=0.9$. A path diagram of the created model is shown in Fig.~\ref{fig:sem-after}.

% \begin{figure}[tbp]
%     \centering
%     \includegraphics[width=86mm]{figure/DSEM-Path_before.pdf}
%     \caption{Static path diagram by exploratory design.}
%     \label{fig:sem-before}
% \end{figure}

In the figure, the nodes enclosed in squares represent observable variables that can be directly observed, while the circular nodes represent latent variables that cannot be directly observed. The edges drawn between the nodes represent causal relationships between the variables. Additionally, green, blue and pink nodes stand for variables at $t-1$, $t$ and $t+1$, respectively.

Real values by the edges  indicate path coefficients in the first experiment and real values in the brackets indicate them in the second experiment. For the analysis of the model, Mplus\footnote{www.statmodel.com} version 8.8 was used, and Bayesian estimation was used to estimate path coefficients.  

The variables and their respective ranges are as shown in the following. AI and human performance means the probability of success for each task. Trust in AI is defined as the human-estimated value of the AI's performance. $AIP$ and $HP$ are the performance of the AI (task success probability of AI) [0, 1] and performance of the human (task success probability of the human) [0, 1]. Also, $E(AIP)$ is the $AIP$ estimated by a human [0, 1], which is the \emph{trust} of a human in AI.
%\begin{equation}
%    \mathrm{E(AIP)} = Estimated(\mathrm{AIP})
%\end{equation}

$Over/Under~Trust$ is used to determine over- or under-trust, where ``$-1$" represents ``under-trust," ``0" represents appropriately calibrated trust, and ``1" represents ``over-trust." $Reliance$ indicates whether the user performed the task themselves, ``0," or delegated the task execution to the AI, ``1." $Calibration\ cue$ indicates whether there was no trust calibration cue, ``0," or a trust calibration cue was presented to the user, ``1." In this model, the edges are drawn, excluding duplicates in the next time step, on the basis of various insights from previous work and our intuition. Note that this model can \emph{directly} determine over/under~trust without the equation for reliance in~\cite{okamura2020adaptive} by introducing the variable $Over/Under~Trust$. We consider this to be the originality of our work. Furthermore, this $Over/Under~Trust$ can be utilized to efficiently and precisely prevent over/under-trust in our future work. 

\section{Experiments}
\subsection{Exp-1 Predicting Trust in Object Recognition with Drone Simulator}
The path coefficients estimated for this experiment are the values in brackets in Fig.~\ref{fig:sem-after}. This model is the result of applying Step~2 optimization of III A with $\eta = 15$ (the maximum length of the time-series data). 
The edges shown in solid lines represent causal relationships that have a positive effect, while the edges shown in dash-dotted lines represent causal relationships that have a negative effect.
The combination of $E(AIP)$ input to the model as a past time series was selected to be $E(AIP)_{(t, t-1)}$ as a result of the model estimation for each combination of all subsets of time $T$.
The combination of $E(AIP)$ corresponding to the combination of time $T$ was selected to be ${t, t-1}$, which is the combination that makes the AIC the smallest among all combinations, after estimating the path coefficients of the model corresponding to all subsets of time $T$.

\begin{figure}[tbp]
    \centering
    \includegraphics[width=80mm]{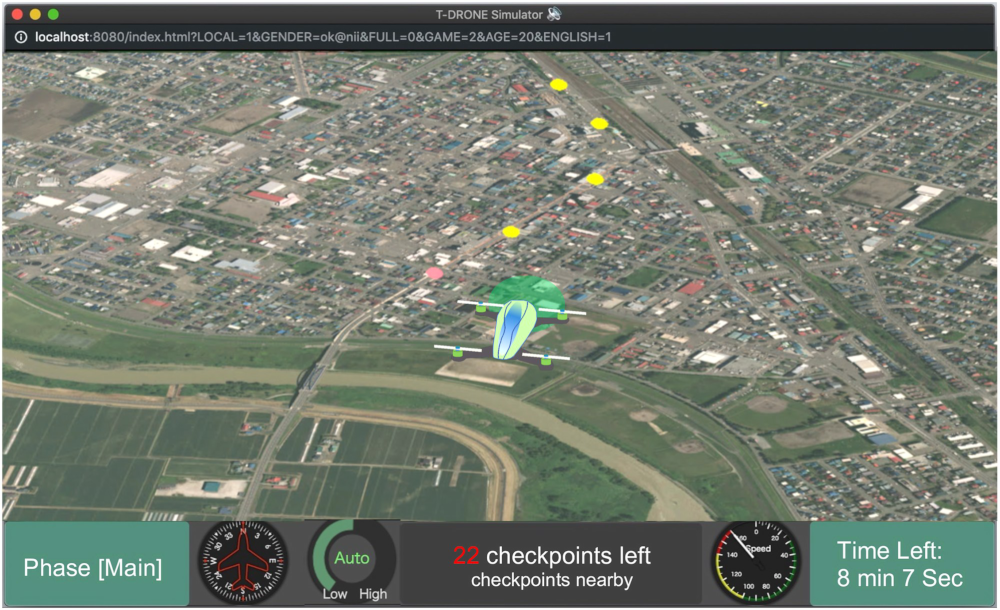}
    \caption{Dron-simulator for human-AI cooperative object recognition~\cite{okamura2020adaptive}.}
    \label{fig:dron-simulator}
\end{figure}

For the estimation of the model, we used the experimental data from a cooperative decision-making task involving image recognition on a drone simulator by~\cite{okamura2020adaptive}. These data were acquired from crowd sourcing-based online experiments with 194 participants, in which humans and AI cooperatively recognize pothole on roads at 30 checkpoints. The snapshot of the simulator is shown in Fig.~\ref{fig:dron-simulator} and these date are completely discrete. The data is consisting two phases: high-performance AI of object recognition in the first 15 checkpoints and low-performance AI in the remaining 15 checkpoints to cause over-trust. Also, adaptive trust calibration was employed and trust calibration cues were expressed to a human when the over-trust was detected. As a result, the data of 96 participants include trust calibration with cues and that of other 96 participants do not include them. For more detail information on the data, see~\cite{okamura2020adaptive}. 

The estimated model allows us to infer the following qualitative causal relationship  from coefficients of each edge:
\begin{itemize}
    \item \emph{Cognitive trust} has positive causality from \emph{AI performance}.
    \item \emph{Over/under trust} has negative causality from \emph{AI performance} and positive causality from \emph{human performance} and \emph{cognitive trust}.
    \item \emph{Cognitive trust} has positive causality from \emph{calibration cue}, while \emph{over/under trust} has negative causality from \emph{calibration cue}.
    \item \emph{Reliance} has positive causality from \emph{cognitive trust}.
    \item \emph{Over/under trust} and \emph{reliance} have positive causality.
\end{itemize}

\subsubsection{Results of Predicting Over-trust}

The results of over-trust prediction at the next time step using the trust prediction model with our proposed method (PM) are shown below. These results are from an analysis of over-trust prediction at the next time step predicted from the current time observation value by each model. To verify the prediction results, we used the experimental data by~\cite{okamura2020adaptive} used for model estimation.

The accuracy (ACC) and root mean squared error (RMSE) of each model are as shown in Table~\ref{tab:result-Drone}. In the prediction with PM, the ACC was 90.0\%, and the RMSE was 0.28.

\begin{table}[tbp]
    \centering
    \caption{Experimental results with drone simulator task.}
    \label{tab:result-Drone}
    \begin{tabular}{lcc}
        \hline
         & ACC & RMSE \\
         & Avg(S.D.) & Avg(S.D.) \\
        \hline \hline
        PM & 0.90(0.05) & 0.28(0.14) \\
        AR(1) & 0.59(0.19) & 0.51(0.19) \\
        ARMA(1,1) & 0.57(0.19) & 0.51(0.19) \\
        SARIMA(1,0,1)[15] & 0.57(0.19) & 0.51(0.19) \\
         \hline
    \end{tabular}
\end{table}

\begin{figure}[tbp]
    \centering
    \includegraphics[width=82mm]{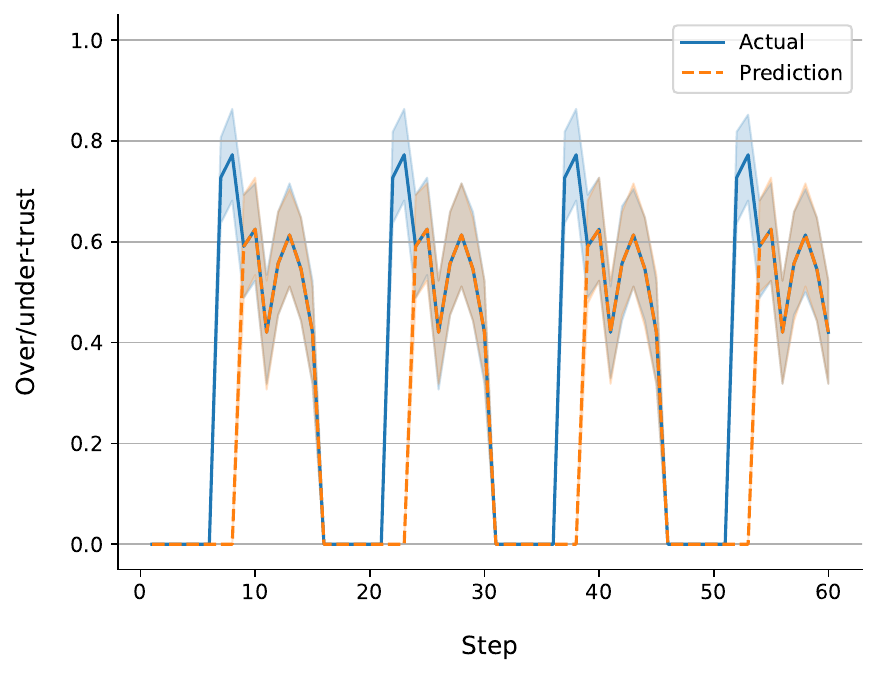}
    \caption{Prediction results of over-trust by DSEM. The blue solid line represents the actual proportion of over-trust, and the orange dashed line represents the predicted proportion.}
    \label{fig:Result_Fittings_Okamura}
\end{figure}

Proportion of users who are actually over-trust and the users who were predicted to be over-trust, step by step are shown in Fig. \ref{fig:Result_Fittings_Okamura}.
The blue solid line represents the proportion of users who actually fell into over-trust, and the orange dashed line represents the proportion of users who were predicted.
Data augmentation was performed using Okamura’s experimental data~\cite{okamura2020adaptive} for path analysis and verification of prediction accuracy by DSEM, resulting in a periodic fluctuation of 15 steps.

\begin{figure}[tbp]
    \centering
    \includegraphics[width=72mm]{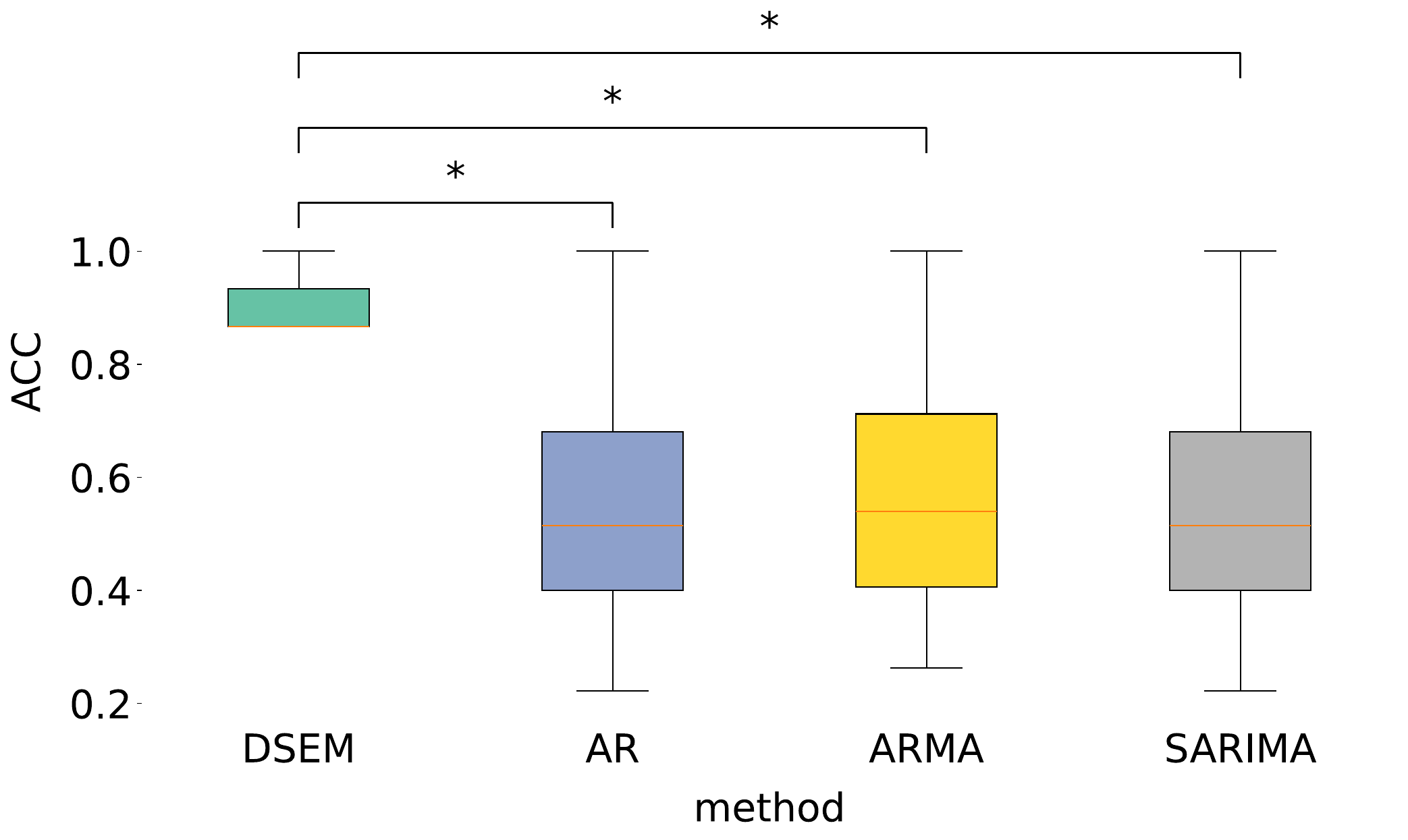}
    \caption{Results of prediction accuracy in a drone simulator task.}
    \label{fig:result-acc_Drone}
\end{figure}

The results of a one-way ANOVA statistical test with multiple comparisons on the accuracy of the prediction by PM and the conventional methods are shown in Fig.~\ref{fig:result-acc_Drone}.
The significance level was set to \(\alpha = 0.05\). As a result, there was a significant difference between PM and all of the base-line methods (autoregression model (AR), autoregressive moving average model (ARMA), seasonal autoregressive moving average model (SARIMA)). This result means our proposed method completely out performed the base-line methods. In all the experiments of this work. the hyper parameters of base-line methods were adequately set based on domain knowledge. Also, VAR (vector autoregression) was not used as a base-line method because AR needs domain knowledge to prepare input/output vectors. 

Also, the precision of the over-trust prediction by DSEM is \(1.00\), and the recall is \(0.72\).

\subsection{Exp-2 Predicting on Autonomous Driving Simulator}

Next, we predicted over/under-trust using an autonomous driving simulator as a more \emph{continuous}-time task in contrast of the previous \emph{discrete}-time drone simulator. The task involved playing video clips from a car-mounted camera on the web and having a human intervene during playback. The video played as an onboard video of an autonomous vehicle was from the BBD100K driving dataset~\cite{bdd100k}. The users were told that the video being played was filmed by an autonomous vehicle. The users played the video on a web browser and indicated their intention to intervene by pressing the space bar on the keyboard when they felt danger while driving. During the experiment, users could continue to monitor the video being played and could intervene as necessary when they felt danger.

The video was played in 22 scenes, with the AI driving at high performance in the first seven scenes, the performance dropping in the next nine, and the AI's performance increasing again in the final six scenes. In the middle nine scenes, if no intervention was made when the AI's performance was lower than that of a human, it was considered to indicate over-trust. Interventions are recorded as one step within a 10-second window, and a total of four steps are recorded within one scene.

\begin{figure}[tbp]
    \centering
    \includegraphics[width=86mm]{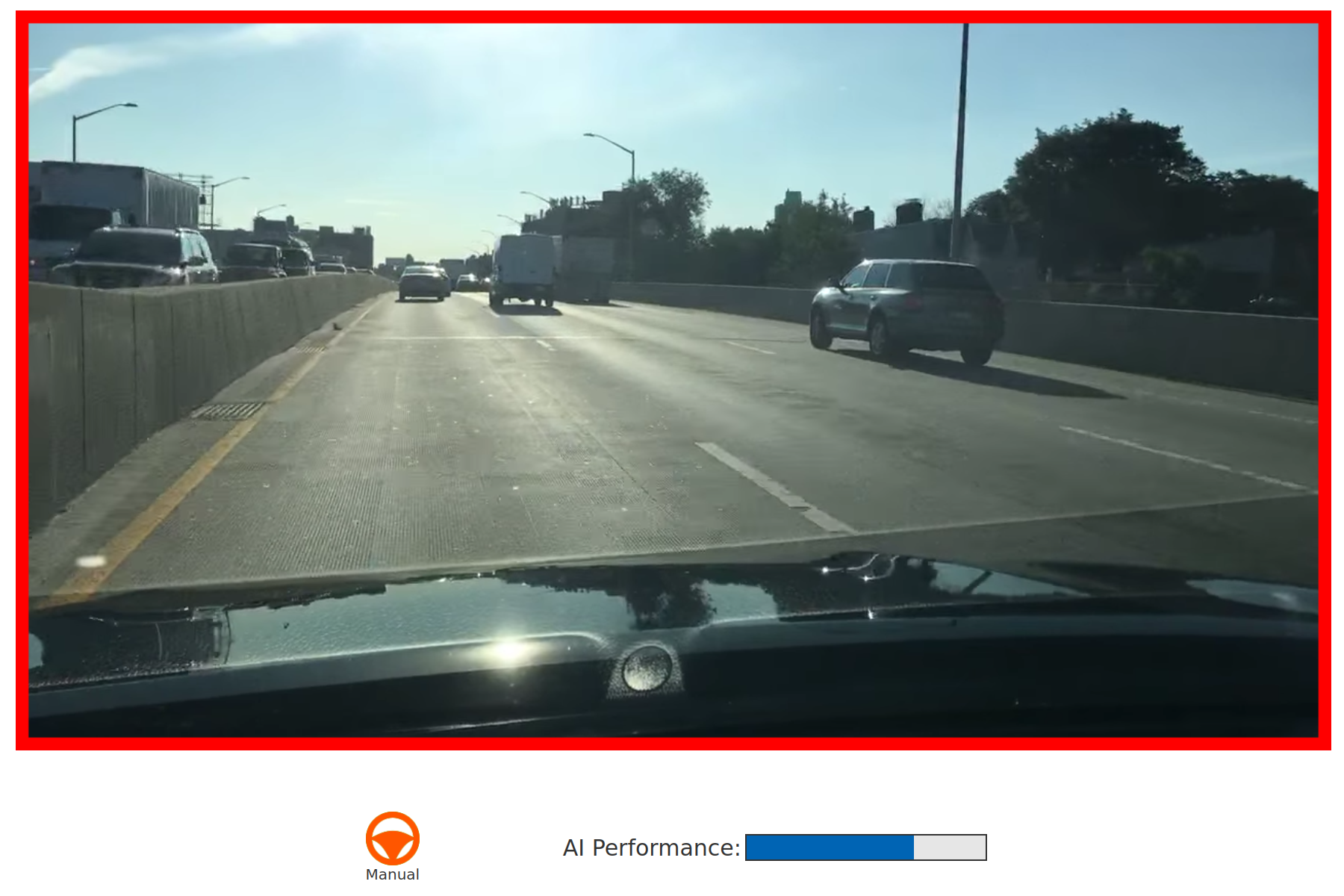}
    \caption{Screen shot of drive simulation while a user was intervening.}
    \label{fig:WebTask_Driving_manu}
\end{figure}

The web-based autonomous driving simulator used in the experiment is shown in Fig.~\ref{fig:WebTask_Driving_manu}.
% Fig.~\ref{fig:WebTask_Driving_auto} shows the situation where autonomous driving is beier intervenes, the screen changes to the one shown red box line as Fig.~\ref{fig:WebTask_Driving_manu}.
When intervention occurs, a red frame appears over the video. %, as shown in the Fig.~\ref{fig:WebTask_Driving_manu}.
While autonomous driving, it becomes a green frame.
A color of the handle icon displayed below the video also changes from green to red when user intervention. Additionally, the capability of AI is directly indicated at the bottom in Fig.~\ref{fig:WebTask_Driving_manu} even though effective representation have been studied~\cite{Chakravarthi2024social}. 

%\begin{figure}[tbp]
%    \centering
%    \includegraphics[width=86mm]{figure/WebTask_20240226_IROS_auto.png}
%    \caption{Screen shot while autonomous driving.}
%    \label{fig:WebTask_Driving_auto}
%\end{figure}

50 participants were recruited for 100 JPY through Yahoo! Japan Crowdsourcing\footnote{https://crowdsourcing.yahoo.co.jp/}. 49 participants (noisy data elimination of two participants) completed the task (11 female, 38 male; aged: 22-66, M = 46.5, S.D. = 9.97).

\subsubsection{Results of Predicting Over/Under-trust}

The results of over-trust prediction at the next time step using the trust prediction model with PM are shown below.

The ACC and RMSE of each model are as shown in Table \ref{tab:result-AV}. In the prediction with PM, the ACC was 97.8\%, and the RMSE was 0.14.

\begin{table}[htbp]
    \centering
    \caption{Experimental results in human-AI cooperative driving task.}
    \label{tab:result-AV}
    \begin{tabular}{lcc}
        \hline
        & ACC & RMSE \\
        & Avg(S.D.) & Avg(S.D.) \\
        \hline \hline
        PM & 0.98(0.01) & 0.14(0.04) \\
        AR(1) & 0.83(0.08) & 0.20(0.10) \\
        ARMA(1,1) & 0.85(0.06) & 0.18(0.08) \\
        SARIMA(1,0,1)[4] & 0.85(0.06) & 0.18(0.08) \\
         \hline
    \end{tabular}
\end{table}

\begin{figure}[tbp]
    \centering
    \includegraphics[width=82mm]{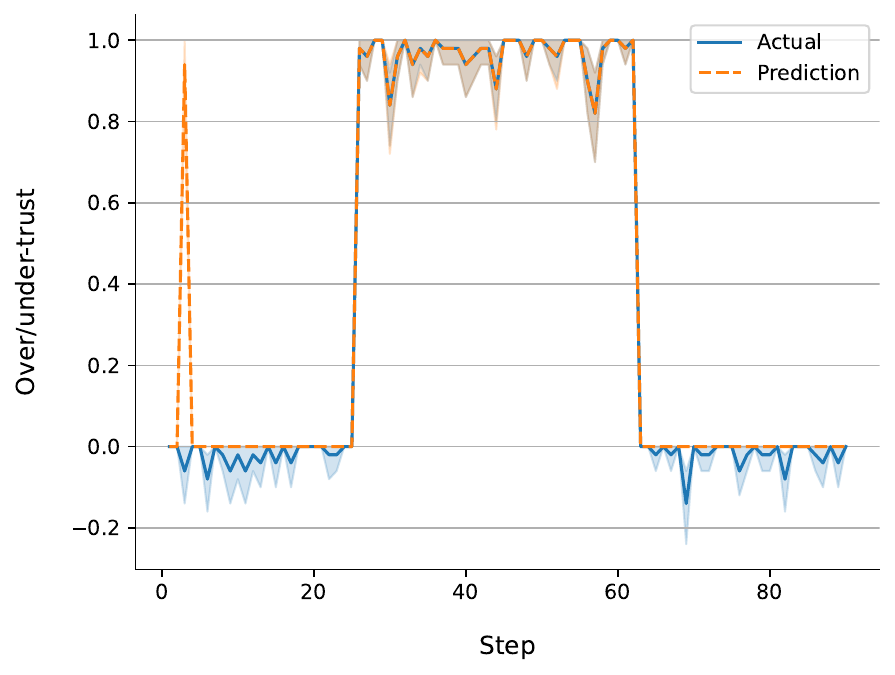}
    \caption{Prediction results of over/under-trust by DSEM. The blue solid line represents the actual proportion of over/under-trust, and the orange dashed line represents the predicted proportion. Positive values represent over-trust, and negative values represent under-trust.}
    \label{fig:Result_Fittings_WebTask}
\end{figure}

The proportion of users who actually engaged in over/under-trust and the users who were predicted to do so, step by step are shown in Fig. \ref{fig:Result_Fittings_WebTask}.
The blue solid line represents the proportion of users who actually fell into over/under-trust, and the orange dashed line represents the users predicted to do so.
Over-trust is shown as positive values and under-trust as negative values.

\begin{figure}[htbp]
    \centering
    \includegraphics[width=72mm]{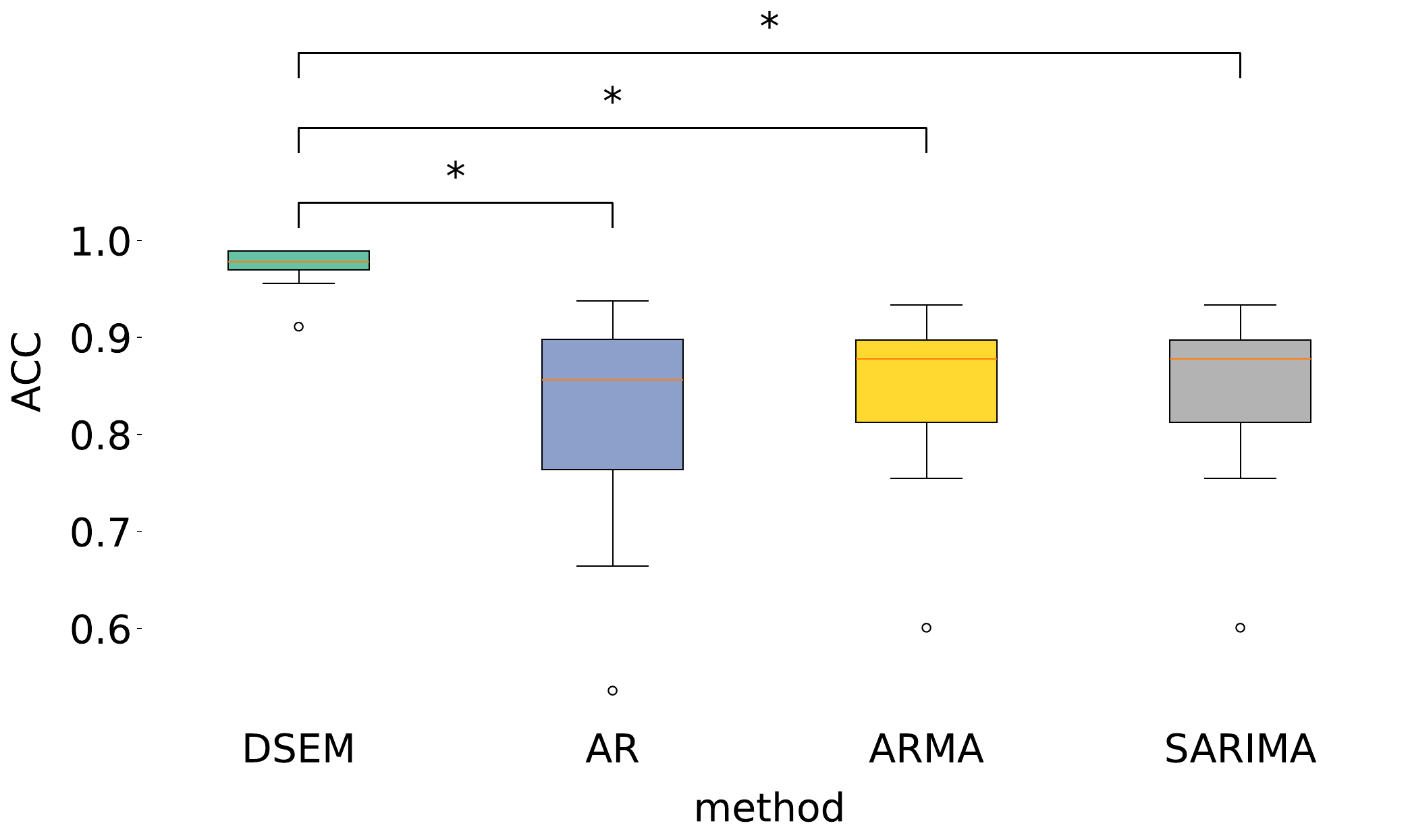}
    \caption{Result of prediction accuracy in a drive simulator task.}
    \label{fig:result-acc_AV}
\end{figure}

The results of a one-way ANOVA statistical test with multiple comparisons on the accuracy of the prediction by PM and the conventional methods are shown in Fig.~\ref{fig:result-acc_AV}.
The significance level was set to \(\alpha = 0.05\). As a result, there was a significant difference between PM and all of the baseline methods (AR, ARMA, SARIMA). This means that our proposed method completely outperformed the baseline methods.

%%%%%%%%%%%%%%%%%%%%%%%%%%%%%%%%%%%%%%%%%%%%%%%%%%%%%%%%%%%%%%%%%%%%%%%%%%%%%%%%
\section{Discussions}

\subsection{Algorithm for Preventing Over/Under-trust}
First, in this study, we focused on over-trust, which poses a bigger problem in particular environments when fallen into than distrust, and made predictions of over-trust. However, of course, the model we built is adaptable to both over-trust and under-trust predictions.

We are currently developing an algorithm for prevention over/under trust with our proposed trust dynamics prediction in this work. The algorithm's basic policy is very simple, that is, to express trust calibration cues to a human just when the over/under-trust is predicted to occur in the next time step.

However, if a human does not react to these cues, what should the AI do? Repeatedly express the cues until the human executes trust calibration? This is not a simple problem so we need to develop an algorithm for preventing over/under-trust through the design of calibration cues. For effective cues, we should carefully design promising cues and conduct experiments to evaluate them. This is our future work. 

 \subsection{Comparison with DNN-based Approaches}
 In the experimental comparison with conventional and baseline systems, we did not include deep neural network (DNN)-based time-series prediction including Transformer~\cite{vaswani2017attention} and LSTM~\cite{Gers2000learning} as state-of-the-arts methods. Our reasons for not utilizing them are because the task properties like snapshots (captured images) of a task simulator, levels of task difficulty, error significance etc., are hard to described and introduced to a SEM framework as observed variables of high-dimensional vectors. In contrast, DNN-based prediction can easily and fully utilize such task properties with embedded vectors as input.

 Thus, it is difficult to prepare the same input for both our proposed method and DNN-based time-series prediction in a fair way. However, we are trying to develop both DNN-based prediction without task description as its input and our proposed methods with task descriptions using high dimensional vectors as observed variables. 

\subsection{Explainability, Interpretability, and Utility of Our Proposed Method}
We think that the explainability and ease of interpretability~\cite{Molnar2023interpretable} of our proposed method can be guaranteed because the prediction models can be described with path diagrams as directed graphs. However, explainability and interpretability were not confirmed in the experiments. Thus, we need to conduct experiments with participants to confirm them. This is also our future work. 

We can utilize the same (static) path diagrams in both experiments in human-AI cooperative object recognition and driving. However, the design of path diagrams is basically dependent on the task domains. Clarifying general and common path diagrams in various task domains, and the coverage of the proposed method are also open problems. 

\subsection{Limitation and Coverage of Our Proposed Method}
Our proposed method has significant limitations. First, the SEM-based approach needs human knowledge because it utilizes an exploratory method with human intuition. This might be a hard limitation depending on the task domain. 

Furthermore, there is no guarantee that precise prediction models will be constructed. Last, our method basically includes a brute-force search with a high computational cost for optimal partial path diagrams. In practice, we can restrict the search space with $\tau$. We are investigating more sophisticated combinatorial optimization algorithms.

We need to discuss the coverage of our proposed approach with DSEM to apply it to other domains. Basically, we consider our approach to be applicable to any domain in which designers have rich knowledge on factors influencing target variables regardless of prediction accuracy. 

Thus, we plan to apply this approach to human-robot interaction and trustworthy AI including the prevention of human abuse of robots~\cite{Ravishankar2024} and robotic trust repair~\cite{Rogers2024}. In particular, for trust repair, we will develop special trust repair cues~\cite{Rogers2024}. 

\section{Conclusion}

In this paper, we proposed a novel method for constructing a prediction model of trust dynamics toward AI in human-AI cooperative decision-making. In our method, first exploratory design is done, and a static path diagram is obtained; then optimization is applied to time-series path diagrams. In this framework, directly predict over/under trust without monitoring the execution of human rational behaviors is quite original and important for preventing over/under-trust. Another advantage of our proposed method is its high explainability due to the path structure. We applied this proposed method to two different task domains involving human-AI cooperative object recognition and autonomous driving. In both domains, we confirmed that our proposed method could outperform conventional methods including AR, ARMA and Seasonal ARMA. 

\bibliographystyle{IEEEtran}
\bibliography{bibliography}

\end{document}